\documentclass[twocolumn,aps,showpacs]{revtex4}

\usepackage{epsfig}

\usepackage{placeins}
\usepackage{lineno}
\begin{document}

\title{Deep sub-threshold \boldmath{$\phi$} production in Au+Au collisions}

\author{J.~Adamczewski-Musch$^{4}$, O.~Arnold$^{10,9}$, C.~Behnke$^{8}$, A.~Belounnas$^{15}$,
	A.~Belyaev$^{7}$, J.C.~Berger-Chen$^{10,9}$, J.~Biernat$^{3}$, A.~Blanco$^{2}$, C.~Blume$^{8}$,
	M.~B\"{o}hmer$^{10}$
% ...
}

\affiliation{
	(HADES collaboration) \\\mbox{$^{1}$Institute of Physics, Slovak Academy of Sciences, 84228~Bratislava, Slovakia}\\
	\mbox{$^{2}$LIP-Laborat\'{o}rio de Instrumenta\c{c}\~{a}o e F\'{\i}sica Experimental de Part\'{\i}culas , 3004-516~Coimbra, Portugal}\\
	\mbox{$^{3}$Smoluchowski Institute of Physics, Jagiellonian University of Cracow, 30-059~Krak\'{o}w, Poland}\\
	\mbox{$^{4}$GSI Helmholtzzentrum f\"{u}r Schwerionenforschung GmbH, 64291~Darmstadt, Germany}\\
	\mbox{$^{5}$Technische Universit\"{a}t Darmstadt, 64289~Darmstadt, Germany}\\
	\mbox{$^{6}$Institut f\"{u}r Strahlenphysik, Helmholtz-Zentrum Dresden-Rossendorf, 01314~Dresden, Germany}\\
	\mbox{$^{7}$Joint Institute of Nuclear Research, 141980~Dubna, Russia}\\
	\mbox{$^{8}$Institut f\"{u}r Kernphysik, Goethe-Universit\"{a}t, 60438 ~Frankfurt, Germany}\\
	\mbox{$^{9}$Excellence Cluster 'Origin and Structure of the Universe' , 85748~Garching, Germany}\\
	\mbox{$^{10}$Physik Department E62, Technische Universit\"{a}t M\"{u}nchen, 85748~Garching, Germany}\\
	\mbox{$^{11}$II.Physikalisches Institut, Justus Liebig Universit\"{a}t Giessen, 35392~Giessen, Germany}\\
	\mbox{$^{12}$Institute for Nuclear Research, Russian Academy of Science, 117312~Moscow, Russia}\\
	\mbox{$^{13}$Institute of Theoretical and Experimental Physics, 117218~Moscow, Russia}\\
	\mbox{$^{14}$Department of Physics, University of Cyprus, 1678~Nicosia, Cyprus}\\
	\mbox{$^{15}$Institut de Physique Nucl\'{e}aire, CNRS-IN2P3, Univ. Paris-Sud, Universit\'{e} Paris-Saclay, F-91406~Orsay Cedex, France}\\
	\mbox{$^{16}$Nuclear Physics Institute, The Czech Academy of Sciences, 25068~Rez, Czech Republic}\\
	\mbox{$^{17}$LabCAF. F. F\'{\i}sica, Univ. de Santiago de Compostela, 15706~Santiago de Compostela, Spain}\\
	\mbox{$^{a}$ also at ISEC Coimbra, ~Coimbra, Portugal}\\
	\mbox{$^{b}$ also at ExtreMe Matter Institute EMMI, 64291~Darmstadt, Germany}\\
	\mbox{$^{c}$ also at Technische Universit\"{a}t Dresden, 01062~Dresden, Germany}\\
	\mbox{$^{d}$ also at Moscow Engineering Physics Institute (State University), 115409~Moscow, Russia}\\
	\mbox{$^{e}$ also at Frederick University, 1036~Nicosia, Cyprus}\\
	\mbox{$^{f}$ also at Dipartimento di Fisica and INFN, Universit\`{a} di Torino, 10125~Torino, Italy}\\
	\mbox{$^{\dag}$ deceased}\\
}  
\date{\today}
\begin{abstract}
We present data on charged kaons ($K^\pm$) and $\phi$ mesons in Au(1.23A GeV)+Au collisions. It is the first simultaneous measurement of $K^-$ and $\phi$ mesons in central heavy-ion collisions below a kinetic beam energy of 10A~GeV. The $\phi / K^-$ multiplicity ratio is found to be surprisingly high with a value of $0.52 \pm 0.16$ and shows no dependence on the centrality of the collision. Consequently, the different slopes of the $K^+$ and $K^-$ transverse-mass spectra can be explained solely by feed-down, which substantially softens the spectra of $K^{-}$ mesons. Hence, in contrast to the commonly adapted argumentation in literature, the different slopes do not  necessarily imply diverging freeze-out temperatures of $K^+$ and $K^-$ mesons caused by different couplings to baryons.
\end{abstract}
\maketitle
Until now, hadron properties and interactions at high baryon densities - as reached in relativistic heavy-ion collisions (HICs) - cannot be addressed directly by ab-initio QCD calculations and thus have to be modeled using effective Lagrangians. Both the equation of state and the kinetic description of the HIC dynamics provide severe challenges with far reaching implications for astrophysical objects, e.g. for neutron star structure and merger dynamics \cite{Nelson,Brown93,Li97,Hanauske:2016gia}.

Strangeness-carrying excitations - with a focus on kaons in the energy regime below 10 A GeV - are considered suitable probes of the properties of compressed nuclear matter and the related collision dynamics \cite{Hartnack11}. In strong interaction processes one observes strangeness production as simultaneous appearance of a 
$s \bar s$ pair, either as a strangeness neutral bound state like the $\phi$ meson or with subsequent redistribution to baryons and mesons (associated production). Arguments based on the OZI rule disfavor the production of the bound $\phi = s \bar s$ state \cite{OZI,shor}. Key quantities of strange mesons are determined by their spectral function related to their so-called nuclear potentials. Various approaches \cite{Lee94,Schaffner97,Lutz94,Koch94,Cassing97,Cabrera14} predict a net attractive $K^-$-nucleon (N) potential. However, due to the presence of baryon resonances \cite{oset,HL}, the resulting $K^-$ spectral function may have an intricate shape and, due to the lack of ab-initio approaches, it needs to be controlled by experimental data. 
The first high-quality data on sub-threshold $K^-$ production in HICs have become available in the late 1990s \cite{Laue99,Menzel00,Forster:2003vc,Forster07}. The data revealed a similar rise of $K^+$ and $K^-$ yields with increasing centrality of the collision, and systematically softer $K^-$ spectra compared to the ones of the $K^+$. Comparisons between data and transport models suggested that the $K^-$ decouples from the system later than the $K^+$ due to the large cross section of strangeness exchange reactions, e.g.~$\pi \Lambda \rightarrow N K^-$, which were predicted in \cite{Ko83} as the dominant source for sub-threshold $K^-$ production. This was taken as an explanation for both, the softer spectra of the $K^-$ (due to the later freeze-out) as well as the similar dependence on the system size (coupling of $K^-$ yield to the one of the $K^+$ via the hyperons) \cite{Hartnack01}. A later freeze-out of the $K^-$ compared to the $K^+$ has thus become a paradigm of sub-threshold strangeness production.
 
Attempting to extract the $K^-$-N potential from experiment, most comparisons between $K^-$ data and transport models favor in fact an attractive  potential. However, quantitative statements are difficult, e.g. due to differences between the various models \cite{Cassing03,Fuchs05,Hartnack11,VZF} 
w.r.t.\ the extracted observable distributions.

The $\phi$ mesons as a possible source of $K^-$ mesons at SIS energies is discussed for the first time in \cite{Chung97,Kampfer01}.
Recent data in light collision systems reveal, indeed, that a sizable fraction of about 20\% of the observed $K^-$ yield results from $\phi$ decays \cite{Mangiarotti02,Agakishiev:2009ar,fphi,Gasik,Piasecki16}. 
The observed difference in the slopes of the $K^+$ and $K^-$ spectra can be explained by taking the $K^-$ contribution from $\phi$ decays into account in light collision systems, as those $K^-$ have a substantially softer spectrum \cite{mB}, as confirmed in \cite{fphi,Gasik,Piasecki16}. Several explanations for the large $\phi/K^-$ ratio in light systems have been proposed, based on both macroscopic \cite{Agakishiev:2010rs,Agakishiev:2015bwu} and  microscopic models \cite{Schade09,Steinheimer:2015sha} but without the emergence of a common picture, and the relation to heavy systems remained vague until now. 

In this letter, we present data on charged kaons ($K^\pm$) and $\phi$ mesons in Au+Au collisions at a kinetic beam energy of 1.23A GeV. It is the only simultaneous measurement of $K^-$ and $\phi$ in central heavy-ion collisions below a kinetic beam energy of 10A~GeV. Mesons with strange quark content are produced deeply below their corresponding free nucleon nucleon thresholds with a clear hierarchy in energy deficits of -150 MeV ($K^+$), -450 MeV ($K^-$) and -490 MeV ($\phi$). Hence, the fireball produced in Au+Au collisons at 1.23A GeV is the
ideal environment to study sub-threshold strangeness production.

The High-Acceptance Di-Electron Spectrometer (HADES) \cite{Agakishiev:2009am} is a charged-particle detector located at the GSI Helmholtz Center for Heavy Ion Research in Darmstadt, Germany. It comprises a 6-coil toroidal magnet centered around the beam axis and six identical detection sections located between the coils covering almost the full azimuthal angle. Low-mass Mini-Drift Chambers (MDCs) are the main tracking detectors, while a scintillator hodoscope (TOF) and a Resistive Plate Chamber (RPC) are used for time-of-flight measurements in combination with a diamond start detector located in front of a 15-fold segmented target. The multiplicity trigger is based on the hit multiplicity in the TOF covering a polar angle range between $45^{\circ}$ and $85^{\circ}$.

\begin{figure}
	\begin{center}
		\resizebox{7.4cm}{!}{%
			\includegraphics{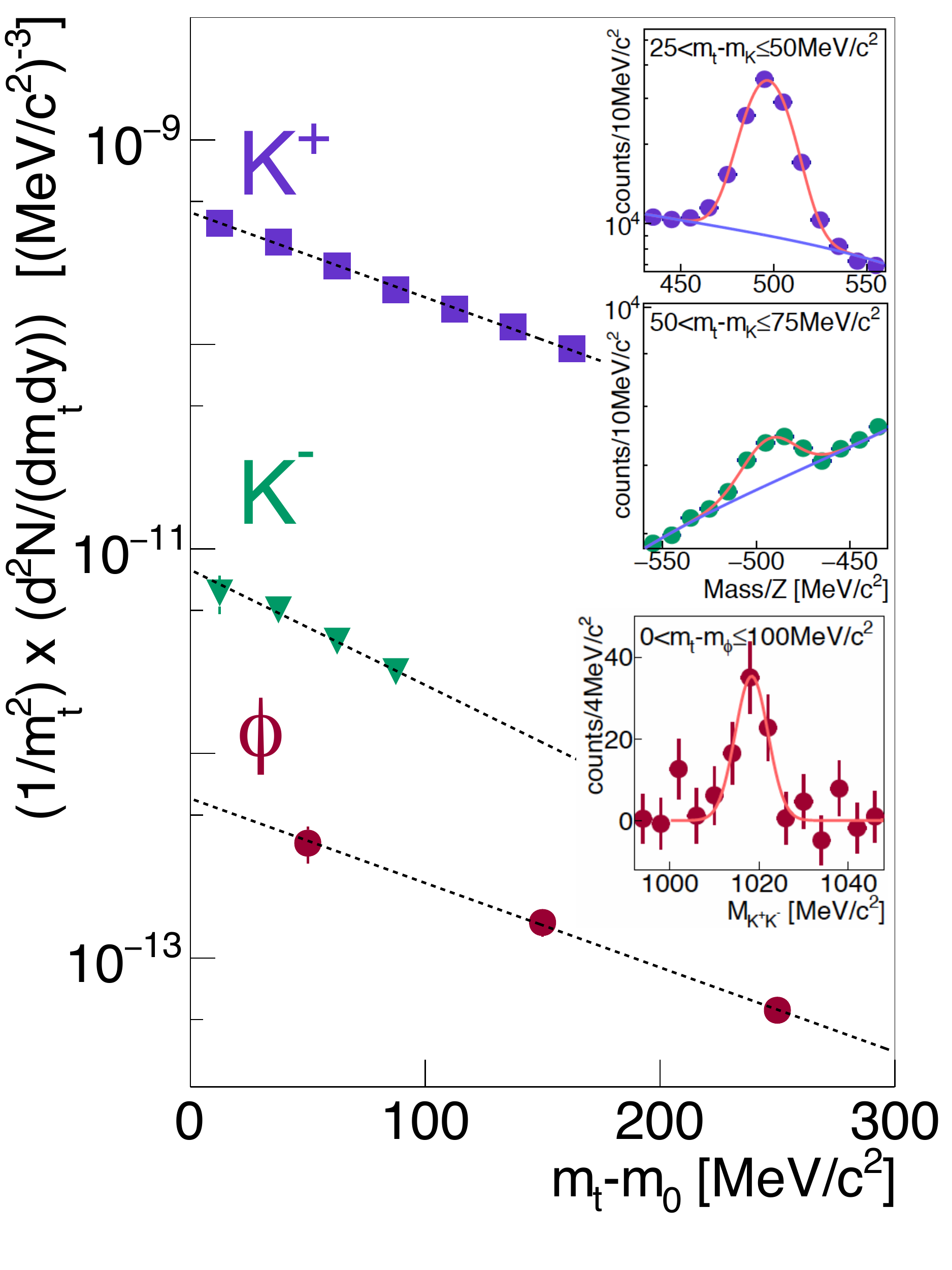}
		}
	\end{center}
	\vspace{-1.2cm}       
	\caption{Left: Acceptance and efficiency corrected transverse-mass spectra around mid-rapidity. The number of counts per event, per transverse mass and per rapidity region, divided by $m_t^2$ together with a fit to the data points according to Eq.~(\ref{bolz}) for the 0-40\% most central events is displayed. Upper right: $K^+$ signal and the corresponding background fit for the region covering mid-rapidity and $m_t-m_0$ between 25 and 50 MeV/$c^2$. The red curve corresponds to the Gaussian part and the blue one to the polynomial part of the combined function used for signal extraction. Middle right: Same as upper one but for $K^-$ and $m_t-m_0$ between 50 and 75 MeV/$c^2$. Lower right: $K^+$$K^-$ invariant mass distribution for the mid-rapidity region and $m_t-m_0$ between 0 and 100 MeV/$c^2$ after subtraction of the background.}
	\label{mt}       
\end{figure}

In total $2.1 \times 10^{9}$ Au+Au events have been collected corresponding to the $40\%$ most central events estimated by elaborated studies using a Glauber model 
\cite{Kardan16}. Charged particle trajectories were reconstructed using the MDC information. The resulting tracks were subject to several selections based on quality parameters delivered by a Runge-Kutta track fitting algorithm. Particle identification is based on the measurements of time-of-flight and track length. Additional separation power for kaons is gained by the energy-loss information from MDC and TOF detectors.
$K^+$ mesons are identified in the center-of-mass rapidity interval of $y_{cm} =-0.65 \cdots +0.25$ in several transverse mass ($m_{t}=\sqrt{p_{t}^2+m_{0}^2}$) bins of 25 MeV/c$^{2}$ width. The underlying background is estimated in an iterative fitting procedure. The fit parameters are observed to show little increase with increasing momentum, exhibiting quantitative agreement with the Monte-Carlo simulation. This procedure allows to obtain the statistical error of the signal and to take into account the quality of the background description of the fit function. Additional variations of the number of parameters, the fit and the integration ranges turned out to be well covered by the error given by the fit. An example of a $K^+$ signal and the corresponding background is displayed in the upper inset of Fig.~\ref{mt} for the region covering mid-rapidity and reduced transverse mass $m_t-m_0$ between 25 and 50 MeV/$c^2$. 
$K^-$ are identified similarly as $K^{+}$ but in a range of $y_{cm}=-0.7$ to $+0.1$. An example of a $K^-$ signal including the background fit is displayed in the middle inset of Fig.~\ref{mt} for the region covering mid-rapidity and $m_t-m_0$ between 50 and 75 MeV/$c^2$.
$\phi$ mesons are identified via their decays into charged kaons. This analysis is done in a rapidity region ranging from $ y_{cm}=-0.5$ to $+0.1$. The combinatorial background is described with the mixed-event technique. For systematic error evaluation on the count rate, the normalization region of the mixed-event and the integration range are varied. The error of the extracted count rate is then defined in the same way as done for $K^\pm$. An example of a $\phi$ signal in the $K^+K^-$ invariant mass after subtraction of the background is displayed on the lower inset of Fig.~\ref{mt} for the mid-rapidity region and $m_t-m_0$ between 0 and 100 MeV/$c^2$.

The raw count rates are corrected in each phase space cell for acceptance and efficiency based on Monte-Carlo and Geant~3 simulations, subject to the same reconstruction and analysis steps as the experimental data.
The average efficiency and acceptance correspond to $\approx$ 0.2 for kaons and about 0.04 for $\phi$ mesons, for details see \cite{heidi}. 
As input for the simulation, thermally distributed $K^{+}$, $\phi$ (T=100 MeV) and $K^{-}$ (T=80 MeV) were embedded into Au+Au collision events generated with the transport code UrQMD \cite{UrQMD} serving as background. 
The systematic bias and uncertainties of the correction are checked based on the more abundant proton and pion tracks as well as on the difference between the different sectors of HADES and are either corrected or taken as systematic error. The total systematic uncertainty on the charged kaon yields within the acceptance corresponds to 4$\%$. The acceptance and efficiency corrected transverse mass spectra of $K^{\pm}$ and $\phi$ for the mid-rapidity bin are presented in Fig.~\ref{mt} in terms of counts per event, per transverse mass and per unit in rapidity, divided by $m_t^2$.. This representation is chosen to ease a comparison with thermal distributions according to
\begin{equation} 
  \label{bolz}
  \frac{1}{m_{t}^{2}} \frac{d^2N}{dm_{t}dy_{cm}} = C(y_{cm}) \,
  \exp \left( -\frac{(m_t-m_0)}{T_B(y_{cm})}  \right) ,
\end{equation} 
where $C(y_{cm})$ is a rapidity dependent normalization, and the slope parameter $T_B$ depends on the rapidity too.
\begin{figure}
 	\begin{center}
 		\resizebox{8.5cm}{!}{%
 			\includegraphics{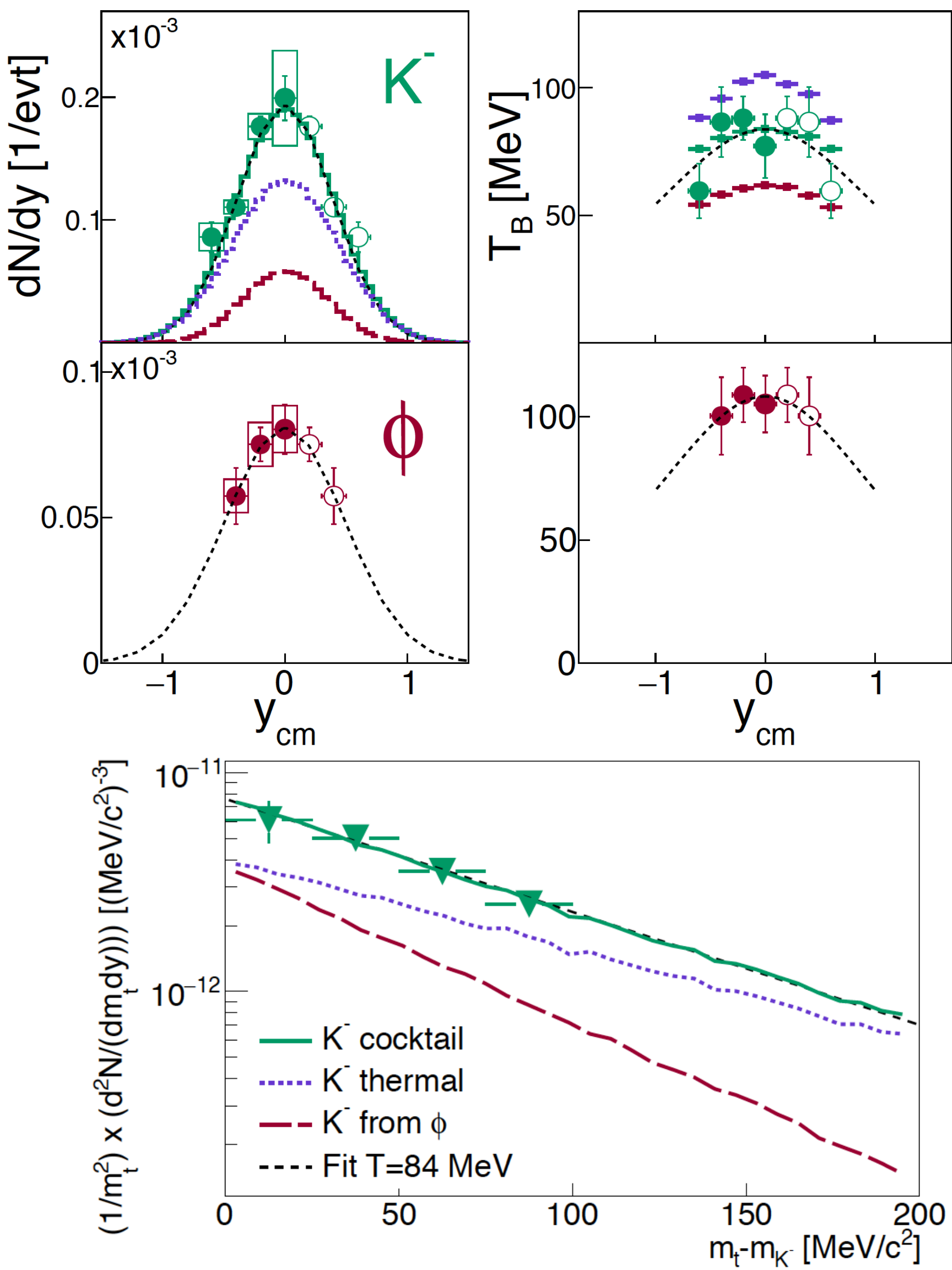}
 		}
 	\end{center}
 	\vspace{-0.6cm}       
 	\caption{Upper left: Rapidity distributions of $K^{\-}$ and $\phi$ and the Gaussian functions (dashed curves) used for extrapolation in $y_{cm}$. Points reflected at mid-rapidity are displayed as open symbols. Upper right: Extracted inverse slope parameters obtained from the Boltzmann fits to the the $m_{t}$ spectra. The distributions are adapted using $T_{B}=\frac{T_{eff}}{\cosh(y_{cm})}$, displayed as dashed curve. Both figures show results for the 0-40\% most central events.
 	In case of the  $K^-$, also the yields and the extracted inverse slopes of the two-component model are displayed: direct thermal (blue), feed-down from $\phi$ decays (red), sum of both (green).
 	Lower panel: $K^-$ transverse-mass spectra around mid-rapidity compared to the different cocktail contributions in the same color code. 
 	}
 	\label{dNdy} 
 	\label{Pluto}      
\end{figure}
Using Eq.~(\ref{bolz}) for an extrapolation in $m_t-m_0$ and integrating the data points, the rapidity density distributions for the different particles are obtained, see Fig.~\ref{dNdy}.  
The uncertainty of the extrapolation is estimated to 1.5$\%$ from the difference between the extrapolation based on Eq.~(\ref{bolz}) and a Siemens-Rasmussen model function including a radial expansion velocity as parameter fixed by using the kinematic distribution of the protons in the same collision system \cite{heidi}. Adding up the different errors quadratically, we find an overall systematic uncertainty on the yield within the covered rapidity range of $\approx 5$\% for charged kaons and of $\approx 10$\% for the $K^{\pm}$ and $\phi$ mesons. Multiplicities are obtained integrating over $y_{cm}$ and using a Gaussian for extrapolation to full phase space. The uncertainty of this extrapolation is estimated based on the relative difference of extrapolated yield obtained for the Gauss and the scaled distribution from UrQMD. The obtained total multiplicities are listed in Tab.~\ref{tab}.
The rapidity distributions and the Gauss functions used for the extrapolation in $y_{cm}$ are displayed on the upper left panel of Fig.~\ref{dNdy} for $K^{\-}$ and $\phi$. The error bars display the statistical error, while the systematic error is indicated by boxes. The extracted inverse slope parameters obtained from the Boltzmann fits to the $m_{t}$ spectra are displayed on the upper right panel of Fig.~\ref{dNdy}. The dependence is fitted using $T_{B}=\frac{T_{eff}}{\cosh(y_{cm})}$ in order to obtain the effective inverse slope $T_{eff}$. While $T_{eff}$ is extracted to $(104 \pm 1_{stat} \pm 1_{sys})$ MeV for $K^+$ and to $(108 \pm 7_{stat})$ MeV for $\phi$ mesons, the obtained value for $K^-$ of $(84 \pm 6_{stat})$ MeV  is significantly lower. This is in line with the previously obtained systematics \cite{Forster07}. The systematic error on the inverse slope of the $K^+$ is obtained by comparison of the spectra extracted in the different sectors separately. In case of the $K^-$ and $\phi$, a similar check gives variations well below the statistical errors and hence are neglected.  
\begin{table}
	\begin{center}
		\begin{tabular}{|c|c|c|c|} \hline
			\textbf{$K^+$} & yield  [10$^{-2}$/evt] & $T_{eff}$ [MeV] \\ \hline \hline
			0 - 40\% & 3.01 $\pm$ 0.03 $\pm$ 0.15 $\pm$ 0.30 &  104 $\pm$ 1 $\pm$ 1 \\ \hline
			0 - 10\% & 5.98 $\pm$ 0.11 $\pm$ 0.30 $\pm$ 0.60 & 110 $\pm$ 1 $\pm$ 1 \\ 
			10 - 20\% & 3.39 $\pm$ 0.05 $\pm$ 0.17 $\pm$ 0.34 &  103 $\pm$ 1 $\pm$ 1 \\ 
			20 - 30\% & 1.88 $\pm$ 0.02 $\pm$ 0.09 $\pm$ 0.19 &  97 $\pm$ 1 $\pm$ 1  \\ 
			30 - 40\% & 1.20 $\pm$ 0.02 $\pm$ 0.06 $\pm$ 0.12 &  91 $\pm$ 1 $\pm$ 1 \\ \hline  \hline 
			
			\textbf{$K^-$} & yield [10$^{-4}$/evt] &  $T_{eff}$ [MeV] \\ \hline \hline
			0 - 40\% & 1.94 $\pm$ 0.09 $\pm$ 0.10 $\pm$ 0.10 &  84 $\pm$ 6 \\ \hline
			0 - 20\% & 3.36 $\pm$ 0.31 $\pm$ 0.17 $\pm$ 0.17 &  84 $\pm$ 7 \\ 
			20 - 40\% & 1.28 $\pm$ 0.11 $\pm$ 0.06 $\pm$ 0.06 &  69 $\pm$ 7 \\ \hline  \hline 
			
			\textbf{$\phi$} & yield [10$^{-4}$/evt] & $T_{eff}$ [MeV] \\ \hline \hline
			0 - 40\% & 0.99 $\pm$ 0.24 $\pm$ 0.10 $\pm$ 0.05  & 108 $\pm$ 7 \\ \hline
			0 - 20\% & 1.55 $\pm$ 0.28 $\pm$ 0.15 $\pm$ 0.11 &  99 $\pm$ 8 \\ 
			20 - 40\% & 0.53 $\pm$ 0.08 $\pm$ 0.05 $\pm$ 0.04 &  91 $\pm$ 7 \\ \hline \hline
			
			& $K^-/K^+$ $\times$ 10$^{3}$ & $\phi/K^-$ \\ \hline \hline
			0 - 40\% & 6.45 $\pm$ 0.9 & 0.52 $\pm$ 0.16 \\ \hline
			0 - 20\% & 7.17 $\pm$ 1.1 & 0.46 $\pm$ 0.12 \\ 
			20 - 40\% & 8.31 $\pm$ 1.3 & 0.44 $\pm$ 0.10 \\ \hline			
		\end{tabular}
	\end{center}
	\vspace{-0.5cm}
	\caption{Multiplicities and effective inverse slopes $T_{eff}$ at mid-rapidity  as well as multiplicity ratios for given centrality classes. The first given error corresponds to the statistical, the second to the systematic error within the rapidity range covered by HADES and the last one to the extrapolation uncertainty to full phase space. If the second or third error is not given, it is found to be well below the statistical error and is hence neglected. The error on the multiplicity ratios corresponds to the quadratic sum of the single error sources.}
	\label{tab}
\end{table}
In addition, the analysis procedure is repeated in four and two centrality classes for the $K^+$ and $K^-$, $\phi$, respectively, which correspond to 10$\%$ (20$\%$) steps in centrality. The results are summarized in Tab.~\ref{tab}. For $T_{eff}$ of $K^-$, we take into account the larger extrapolation in $m_t-m_0$, due to the reduced statistics, by an additional systematic error.

\begin{figure}
	\begin{center}
		\resizebox{5.3cm}{!}{%
			\includegraphics{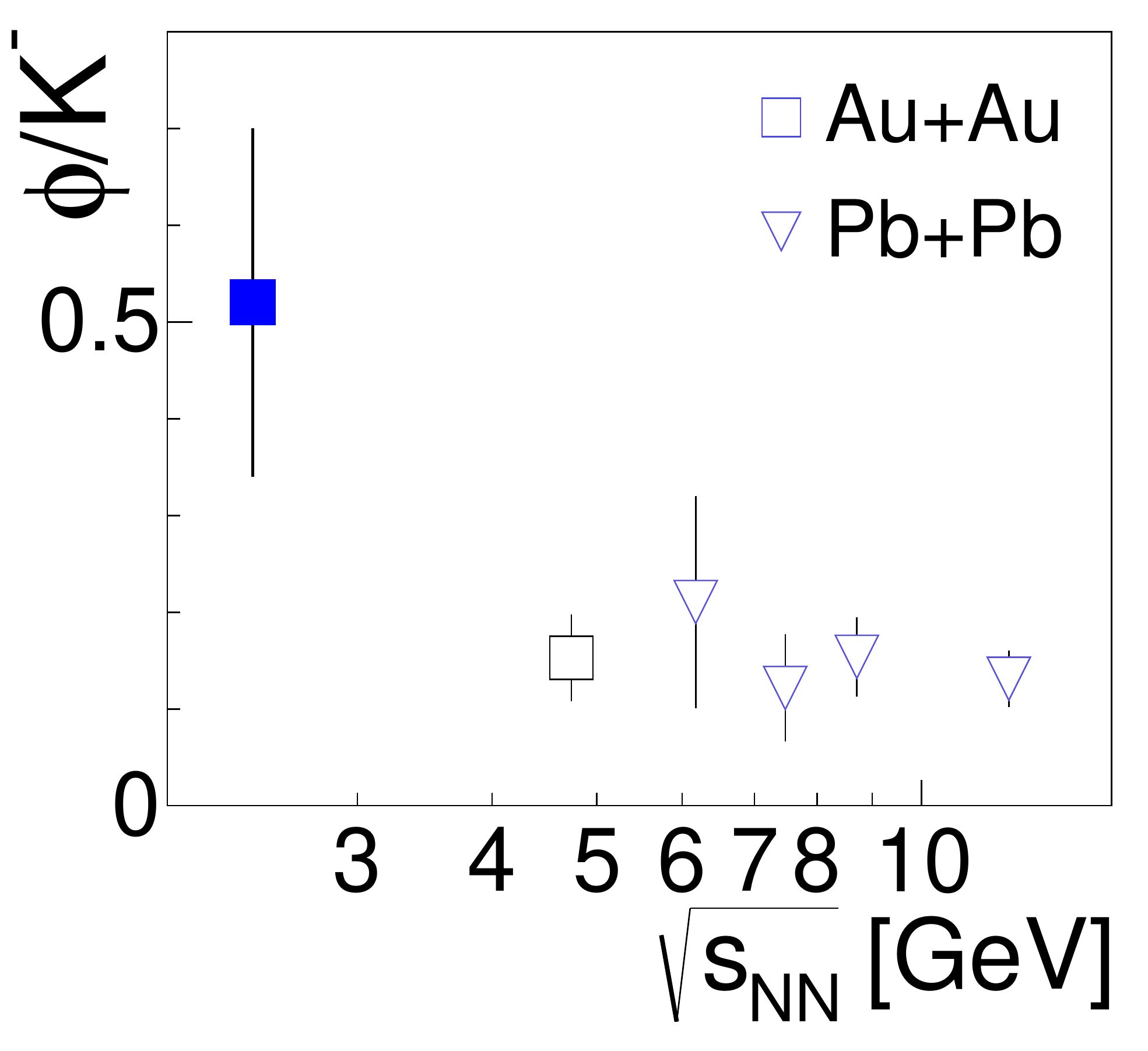}
		}		
	\end{center}
	\vspace{-0.8cm}       
	\caption{
		Multiplicity ratio $\phi$/$K^-$ as function of $\sqrt{s_{NN}}$ for central heavy-ion collisions \cite{E917,NA49}; the HADES data is depicted as filled symbol.}
	\label{KmKP}       
\end{figure}

The hierarchy in energy deficits is reflected in the yields of the three mesons: $K^+$ mesons are found to be two orders of magnitude more abundantly produced than the deep sub-threshold produced $K^-$ and $\phi$ mesons. Due to the similar rise of charged kaon yields with increasing centrality, one can directly compare the extracted ratio $K^-/K^+= (6.45\pm 0.77) \times 10^{-3}$ to values obtained by the KaoS collaboration at higher beam energies and various collision systems \cite{Forster07,Agakishiev:2009ar} without correcting for the different centrality selections. The ratio shows a linear increase with $\sqrt{s_{NN}}$, and our data point is consistent with the extrapolation from higher energies using a linear regression \cite{Agakishiev:2009ar}.  
The yield and slope of the $\phi$ meson have never before been measured in central heavy-ion collisions below a kinetic beam energy of 10A~GeV. The excitation function of the $\phi/K^-$ ratio is depicted in Fig.~\ref{KmKP} as a function of $\sqrt{s_{NN}}$, including data from higher energies \cite{E917,NA49}. Assuming the validity of the previous paradigm of sub-threshold strangeness production presented in the introduction, one expects the $\phi/K^-$ to decrease with decreasing energy, as it becomes increasingly unlikely to accumulate enough energy for $\phi$ production, while $K^-$ can still be produced via strangeness exchange reactions, which should have sufficient time to occur in large systems. However, while the ratio is constant $\approx$ 0.15 for $\sqrt{s_{NN}}\ge$ 4 GeV, our data indicate a strong increase towards low energies: We find a $\phi/K^-$ ratio of $0.52 \pm 0.16$.

This rises questions on the widespread assumption of a small $\phi$ production cross section \cite{shor} due to the OZI rule and shows that indeed correlated kaon production via $\phi$ mesons is a sizable source of $K^-$ ($(26\pm8)\%$ of all $K^-$mesons) in large collision systems at low energies.  
Hence, the feed-down from $\phi$ meson decays can not be neglected in the $K^-$ channel. To investigate the feed-down effect we built a simple $K^-$ cocktail using the event generator Pluto \cite{Frohlich:2009sv}. We generate two static thermal sources, one for direct $K^-$ and one for $\phi$ mesons with temperatures of $T=104$ MeV and $T=108$ MeV, respectively, according to the measured inverse slopes of $K^+$ and $\phi$. Due to the hierarchy in production yields the feed-down on the $K^+$ spectra is negligible. In case of the $K^-$, we scale the two contributions according to the measured $\phi/K^-$ ratio. The different cocktail contributions to the $K^-$ transverse-mass spectra around mid-rapidity are displayed in the lower panel of Fig.~\ref{Pluto}: direct thermal (blue), resulting from $\phi$ decays (red), sum of both (green). It turns out that the $K^-$ resulting from $\phi$ decays have a much softer spectrum and hence substantially "cool" the finally observed spectrum. The sum of both contributions is then fitted using Eq.~(\ref{bolz}) (black) in a similar $m_t-m_0$ range between 0 and 200 MeV/$c^2$ as used for experimental data (green triangles), both displayed in Fig.~\ref{Pluto}. The rapidity dependences of the extracted inverse slopes for different $K^-$ sources are displayed on the right panel in Fig.~\ref{Pluto} with the same color code as above.
The inverse slope of ($84 \pm 5$) MeV agrees with the measurement of $(84\pm 6)$ MeV. The error is obtained by variation of the  $\phi/K^-$ ratio within the given errors. The error on the inverse slope parameter of the experimental spectrum is propagated by making use of the covariance matrix when determining the yields and hence is not varied explicitly. We find the shape of the rapidity distribution to be reproduced as well, displayed together with the $K^-$ data on the left panel of Fig.~\ref{Pluto}, where a comparison of the data to the full cocktail (green curve), the direct (blue curve) and contribution from $\phi$ decays (red curve) is shown. The different slopes of the $K^+$ and $K^-$ transverse-mass spectra can be explained solely by feed-down, which substantially softens the spectra of $K^{-}$ mesons and do hence not imply different freeze-out temperatures of both mesons resulting from their unequal couplings to baryons.

The general understanding of sub-threshold strangeness production is further challenged by the investigation of the centrality dependence of the $\phi/K^-$ and $K^-/K^+$ ratios. Within the previous paradigm, one also expects the relative yields to show different scalings with the system size as the $K^-$ yield is coupled to the one of $K^+$ via strangeness exchange reactions, while no such reactions are present in case of the $\phi$ meson (note that the average amount of produced strange quark pairs is at the order of $10^{-2}$ per event). Hence, a significantly higher amount of energy must be accumulated before $\phi$ meson production can occur and therefore a stronger scaling with increasing centrality is expected, than for the charged kaons. However, both the $\phi/K^-$ and the $K^-/K^+$ ratios extracted within the two centrality classes do not show any increase towards central events, see Tab.~\ref{tab}. This implies that energy is much more easily redistributed in the created fireball than previously assumed, suggesting a universal scaling of produced strangeness with increasing system size.

Several ongoing calculations promise to improve our understanding of sub-threshold strangeness production and will be confronted with the data for more quantitative comparisons in the future: An improved version of the transport code UrQMD \cite{Steinheimer:2015sha} can describe the observed $\phi/K^-$ for energies at low energies by increasing the $\phi$-N coupling via higher baryon resonances, which decay to final states including a $\phi$ meson and act as energy reservoir at the same time. As such decay branches are not directly observed, their branching ratios are tuned to match data on elementary $\phi$ production cross sections \cite{Maeda07}. In support of this, recent data from elementary collisions \cite{ozi_pp} and investigations about the $\phi$ meson self-energy in nuclear matter \cite{Cabrera:2016rnc} show stronger $\phi$-N couplings than expected by the OZI rule. Also statistical models can reproduce the rise of the $\phi/K^-$ ratio when including the so-called strangeness correlation radius $R_c$, which governs the canonical suppression. As the $\phi$ meson has no net-strangeness, it is not affected by strangeness conservation in the reduced volume and therefore not suppressed, while the $K^-$ meson is. This results in an increase of the $\phi$/$K^-$ ratio with decreasing $\sqrt{s_{NN}}$, where the size of $R_c$ determines the strength of the decrease. Also, the similar dependence of charged kaon and $\phi$ production with the centrality of the collision is naturally reproduced as the total amount of produced strangeness increases with decreasing impact parameter of the collision, prior to its redistribution to the different hadron species at freeze-out \cite{kolo}.
In summary, we have presented first combined data on charged kaons ($K^\pm$) and $\phi$ mesons in Au+Au collisions at $\sqrt{s_{NN}}=2.4$ GeV.    
The $\phi / K^-$ ratio shows a increase with decreasing center-of-mass energy $\sqrt{s_{NN}}$ and is found to be $0.52 \pm 0.16$ in our experiment. Hence, with a fraction $(26\pm8)\%$ of $K^-$mesons resulting from $\phi$ meson decays, the latter one turns out to be a sizable source of antikaon production. The different slopes of the transverse $K^+$ and $K^-$ spectra can be fully explained by feed-down. The $\phi/K^-$ ratio is constant as a function of centrality, suggesting a universal scaling of produced strangeness with increasing system size. Both observations have not been taken properly into account by phenomenological models in the past and thus further developments are needed to arrive at firm conclusions on the $K^-$-N potential, which is an important ingredient, e.g. for the description of astrophysical objects, as mentioned in the introduction. Conclusions based on that potential must be tested with regard to their consistency with our findings. An improved understanding of strangeness dynamics in HICs is also a necessary prerequisite for sharpening the science case of experiments at the upcoming large scale facilities as FAIR, NICA, J-PARC and the low energy runs of RHIC. 

The HADES collaboration gratefully acknowledges the support by the grants SIP JUC Cracow (Poland), 2013/10/M/ST2/00042; TU Darmstadt (Germany), VH-NG-823; GU Frankfurt, (Germany), BMBF:05P15RFFCA, HIC for FAIR, ExtreMe Matter Institute EMMI; TU M{\"u}nchen, Garching (Germany), MLL M{\"u}nchen, DFG EClust 153, DFG FAB898/2-1, BmBF 05P15WOFCA; JLU Giessen (Germany), BMBF:05P12RGGHM; IPN, IN2P3/CNRS (France); NPI CAS Rez (Czech Republic), GACR 13-06759S, MSMT LM2015049.

\end{document}